\documentclass[%
 reprint,
superscriptaddress,
 amsmath,amssymb,
 aps,
]{revtex4-2}

\usepackage[utf8]{inputenc}
\usepackage{amsmath}
\usepackage{bm}
\usepackage{graphicx}
\usepackage{amssymb}
\usepackage{times}
\usepackage{color}
\usepackage{float}
\usepackage{multirow}
\usepackage{subcaption}
\usepackage{braket}
\usepackage{hyperref}
\usepackage{chemformula}
\usepackage{makecell}
\usepackage{xfrac} 
\usepackage{threeparttable} 

\begin{document}

\title{Na in Diamond: High Spin Defects Revealed by the ADAQ High-Throughput Computational Database}

\author{Joel Davidsson}
\email{joel.davidsson@liu.se}
\affiliation{Department of Physics, Chemistry and Biology, Link\"oping University, Link\"oping, Sweden}

\author{William Stenlund}
\affiliation{Department of Physics, Chemistry and Biology, Link\"oping
 University, Link\"oping, Sweden}

\author{Abhijith S Parackal}
\affiliation{Department of Physics, Chemistry and Biology, Link\"oping University, Link\"oping, Sweden}

\author{Rickard Armiento}
\affiliation{Department of Physics, Chemistry and Biology, Link\"oping University, Link\"oping, Sweden}

\author{Igor A. Abrikosov}
\affiliation{Department of Physics, Chemistry and Biology, Link\"oping University, Link\"oping, Sweden}

\begin{abstract}

Color centers in diamond are at the forefront of the second quantum revolution.
A handful of defects are in use, and finding ones with all the desired properties for quantum applications is arduous.
By using high-throughput calculations, we screen 21607 defects in diamond and collect the results in the ADAQ database.
Upon exploring this database, we find not only the known defects but also several unexplored defects.
Specifically, defects containing sodium stand out as particularly relevant because of their high spins and predicted improved optical properties compared to the NV center.
Hence, we studied these in detail, employing high-accuracy theoretical calculations. 
The single sodium substitutional (Na$\mathrm{_C}$) has various charge states with spin ranging from 0.5 to 1.5, ZPL in the near-infrared, and a high Debye-Waller factor, making it ideal for biological quantum applications.
The sodium vacancy (NaV) has a ZPL in the visible region and a potential rare spin-2 ground state.
Our results show sodium implantation yields many interesting spin defects that are valuable additions to the arsenal of point defects in diamond studied for quantum applications.

\end{abstract}

\maketitle


\section{Introduction}

Of all point defects in diamond, the NV center~\cite{davies1976,doi:10.1126/science.276.5321.2012,doi:10.1063/1.1507838} stands out as the most studied and used in quantum applications with many ongoing parallel efforts.
It fulfills many of the defect properties listed for various quantum applications~\cite{Wolfowicz2021}.
The NV center is the main defect considered as a computational qubit~\cite{Nizovtsev2005,doi:10.1073/pnas.1003052107} with recent advancements in fault-tolerant operation~\cite{Abobeih2022}.
While more work is needed before it can be realized at large scale~\cite{doi:10.1063/5.0007444}, it has already demonstrated promise as a flying qubit by transmitting quantum information over long distance~\cite{Hensen2015} and multinode network capabilities~\cite{doi:10.1126/science.abg1919,PhysRevX.12.011048}, which goes towards the quantum internet~\cite{Kimble2008,doi:10.1126/science.aam9288}.
It has also been demonstrated as memory for quantum information~\cite{doi:10.1126/science.1131871,PhysRevX.9.031045}.
Recent advances in quantum sensing with NV centers include magnetrometry~\cite{PhysRevLett.128.177401} at extreme conditions~\cite{hilberer2023nv}, nano-scale nuclear magnetic resonance~\cite{Abobeih2019,D2CC01546C,doi:10.1073/pnas.2111607119}, relaxometry~\cite{doi:10.1021/acs.accounts.2c00520}, and biological applications~\cite{doi:10.1126/science.aad8022,doi:10.1073/pnas.2114186119,doi:10.1126/sciadv.aaw7895}.
These are just some of the explored and proposed applications for the NV center.
It is a genuinely versatile defect that is also well understood from the theoretical side~\cite{Doherty_2011,GaliNV} with known properties such as spin-1, many-body structure etc.
However, the NV center does have some drawbacks.
It has a Zero Phonon Line (ZPL) in the visible range (outside the first and second biological window~\cite{Weissleder2001,Smith2009} as well as the telecom region~\cite{gasca2008future}) with a low Debye-Waller factor ($\sim$3.2\%~\cite{Alkauskas_2014}) and is affected by spectral diffusion~\cite{PhysRevLett.110.027401,PhysRevLett.110.167402,PhysRevApplied.18.064011}.

Other defects in diamond improve on these aspects.
The group 14 (Si~\cite{Wang_2006,Neu_2011,Müller2014,doi:10.1126/science.aao0290}, Ge~\cite{Palyanov2015,Iwasaki2015}, Sn~\cite{PhysRevLett.119.253601,Tchernij2017}, and Pb~\cite{PhysRevB.99.075430,DitaliaTchernij2018}) vacancy centers~\cite{Bradac2019} have a wide range of ZPLs from near-infrared to visible with higher Debye-Waller factors, from about 20\% to 70\%~\cite{gergo18}.
Since the dopant sits between two vacancy positions, known as a split-vacancy configuration, these defects have D$\mathrm{_{3d}}$ symmetry, specifically inversion, that suppresses spectral diffusion.
This property, combined with the high Debye-Waller factor, makes these defects ideal for photonics~\cite{doi:10.1126/science.aah6875,PhysRevLett.118.223603,Guo2021}.
They are also considered as spin qubits, and as the ion size increases, the ground state splitting becomes larger, providing coherent spin control as demonstrated for the SnV~\cite{Görlitz2022,PhysRevX.11.041041}.
However, the group 14 vacancy centers also have some drawbacks.
They are spin-\sfrac{1}{2}, one of the defect states are below or just above to the valence band edge, and the ZPLs are in the visible range.

There are a handful of other less studied defects in diamond, mainly other vacancy complexes in a split-vacancy configuration.
Theoretically suggested defects include the group 13 (Al, Ga, In, and Tl) vacancy centers~\cite{PhysRevB.102.195206}.
These centers have spin-1 and ZPLs in the visible region with Debye-Waller factors from about 15\% to 43\%.
Another vacancy defect is the MgV center with spin-\sfrac{1}{2} in the negative charge state with a Debye-Waller factor of 54\%~\cite{Pershin2021}.
This defect is found along with single substitutional Mg during Mg implantation~\cite{Corte2023}.
Yet another vacancy complex is the NiV with spin-\sfrac{1}{2}, ZPL in the near-infrared range, and predicted Debye-Waller factor between 28\% and 74\%~\cite{PhysRevResearch.3.043052}.
Recent experimental study announced the Debye-Waller factor for this defect is 51\%~\cite{morris2023electronic}.
The \ch{Ni_C} (W8) is a related defect with notably spin-\sfrac{3}{2}~\cite{NAZARE2001616,PhysRevB.87.245206}.
This defect is one of the few with spin-\sfrac{3}{2} reported in diamond.
The most studied spin-\sfrac{3}{2} defect is the silicon vacancy in SiC~\cite{vacancypaper}.
The high spin gives unique sensing opportunities for strain and temperature which are universal for defects with spin larger or equal to spin-\sfrac{3}{2}~\cite{PhysRevB.95.081405}.
So far, no spin-2 defect in diamond has been reported.
More diamond defects can be found in Refs.~\onlinecite{Aharonovich_2011,THIERING20201}.
High-throughput experimental search has studied known defects in diamond~\cite{Luhmann_2018}.

Apart from the known defects in diamond, there are many unknown defects~\cite{zaitsev2013optical}.
Notable examples include: ST1 (spin-0, ZPL in the visible range, oxygen related)~\cite{Lee2013,BalasubramanianMetschReddyRogersMansonDohertyJelezko+2019+1993+2002,Lühmann2022}; TR12 (spin 0, ZPL in the visible range, static Jahn–Teller distortion)~\cite{GDavies_1981,Foglszinger2022}; implanted defects with F (ZPL in the visible range, possibly FV center with spin-\sfrac{1}{2} or spin-1)~\cite{DitaliaTchernij2020}, and Xe (ZPL in the ultraviolet range with high Debye Waller around 74\%, possibly XeV center with spin-\sfrac{1}{2} or spin-1)~\cite{SANDSTROM2018182}.

Given all these defects, has the best defect in diamond been found?
To answer this, we turn to theoretical high-throughput methods.
Previous such studies systematically looked at vacancy center complexes~\cite{PhysRevB.72.035214}.
However, this study limits the dopants to p-elements and the defects to vacancy centers.
The large space of point defects in diamond remains unexplored.

In this paper, we screen defects in diamond using the high-throughput framework ADAQ (Automatic Defect Analysis and Qualification)~\cite{ADAQ,adaq_info}, which in turn uses the high-throughput toolkit (\emph{httk})~\cite{armientoDatabaseDrivenHighThroughputCalculations2020}.
The ADAQ software package is a culmination of a series of publications related to SiC~\cite{methodologypaper,vacancypaper,6Hpaper,stackingpaper} that focus on accurately calculating magneto-optical properties for point defects, such as ZPL.
ADAQ has been successfully applied to SiC and CaO, where modified silicon vacancy~\cite{modvac} and $\mathrm{X_{Ca}V_O}$ defects with X=Sb, Bi, and I~\cite{davidsson2023ab} were found.
In this paper, we turn our attention to diamond to investigate single and double defects consisting of vacancies as well as substitutional and interstitial s- and p-elements, however, we do exclude interstitial-interstitial clusters (due to sheer size of them, about 90000 defects).
We include defects that are separated up to 2.6 $\mathrm{\AA}$, which roughly corresponds to second nearest neighbors and we only make double defects with one external dopant.
The result is a total of 21607 defects to consider.
These defects were screened at the PBE DFT level (see Sec.~\ref{sec:methodology} for more on the workflow and calculational details), and the results were stored to the ADAQ database.
The stored properties include: formation energy, charge transition levels, defect levels, defect spin, zero phonon line with radiative lifetime and transition dipole moment (including polarization), $\Delta$Q (relaxation between ground and excited state), and more.

This paper is organized as follows:
Sec.~\ref{sec:res} is divided into two main parts, database searches and spin defects with sodium.
Sec.~\ref{sec:database} describes how the most stable defects are found on the defect hull, and discusses ZPL accuracy for the previously known defects in diamond.
Here, we also narrow down the number of relevant defects for quantum applications by multiple searches that filter out defects with NV-like properties (spin-1 and a bright ZPL, Sec.~\ref{sec:spin-1}), higher spin states (spin-\sfrac{3}{2} Sec.~\ref{sec:spin-3/2} and spin-2 Sec.~\ref{sec:spin-2}), or high Debye-Waller factors (Sec.~\ref{sec:Qdefects}).
When considering all these properties, the sodium substitutional and vacancy center stand out.
Hence, in Sec.~\ref{sec:nac} and Sec.~\ref{sec:nav}, we studied these defects further with higher-order methods (such as HSE DFT calculation) to confirm their properties.

\section{Results}
\label{sec:res}

This section is divided into two main parts: several sections exploring different point defects in diamonds and two more sections examining spin defects with sodium in more detail.

\subsection{Database Search: Intro}
\label{sec:database}

The most stable defects are located on the defect hull---the lowest formation energy per stoichiometry and per Fermi energy~\cite{davidsson2021color,modvac}.
In the ADAQ database, all known defects, consisting of s- and p-elements, mentioned in the introduction are found on the defect hull.
In this section, we discuss these defects and their optical properties compared with experiment and other theoretical calculations.

ADAQ predicts the ZPL of defects with at least one occupied and unoccupied defect state within the band gap.
However, due to the use of the PBE functional, they are systematically underestimated.
For the NV center, ADAQ predicts a ZPL of 1.700 eV (see Table~\ref{tab:spin-1}), whereas the experimental value is 1.945 eV~\cite{davies1976} (a difference of 0.245 eV).
For the group 14 vacancy defects; SiV, GeV, and SnV do not have a ZPL in the database because one of the defect states involved in the transition is below the valence band edge~\cite{gergo18}.
However, as the dopant gets larger, the defect state enters the band gap~\cite{gergo18}.
Hence, the PbV center does have a state in the band gap, which ADAQ finds and predicts a ZPL of 2.122 eV, which is close to the measured 2.384 eV (520 nm)~\cite{PhysRevB.99.075430} (a difference of 0.263 eV).
In general, we find from comparing the ADAQ ZPL predictions from the screening workflow to experimentally measured defects that the ZPLs are underestimated by around 0.25 eV (mean difference).
Similar mean difference and variation are observed for defects in SiC~\cite{methodologypaper,ADAQ,modvac}.

One can also compare the ADAQ ZPL predictions with other theoretical results.
For the group 13 vacancy defects, the ADAQ ZPLs are AlV 1.00 eV, GaV 1.72 eV, InV 1.87 eV, and TlV 2.31 eV (see Table~\ref{tab:spin-1}).
When comparing with the HSE calculations done in Ref.~\onlinecite{PhysRevB.102.195206}, the ZPLs are for GaV 1.82 eV (difference 0.1 eV), InV 2.12 eV (difference 0.25 eV), and TlV 2.84 eV (0.53 eV).
The ZPL for the AlV center is not reported due to numerical convergence issues~\cite{PhysRevB.102.195206}.

Finally, the MgV center is on the defect hull, and ADAQ reports a ZPL of 0.31 eV.
In the screening workflow, ADAQ calculates only one excitation.
This excitation for the MgV defect matches with the theoretical results in Ref.~\onlinecite{Pershin2021}, where the lowest excitation ($\mathrm{^2E_g-^2E_u}$ in the doublet state without Jahn-Teller effect) has an adsorption value of 0.7 eV.

\subsection{Database Search: NV-like Spin-1 Defects}
\label{sec:spin-1}

To look for NV-like defects, we search for spin-1 defects on the defect hull with a ZPL larger than 0.5 eV and a transition dipole moment (TDM) larger than 3 debye.
We find 15 unique defects, see Table~\ref{tab:spin-1}.
Apart from the already discussed substitutional vacancy complexes (the NV center and group 13 vacancies), there are also vacancy centers consisting of Li and Ba.
Both of these show similar properties as the NV center, but with lower ZPL and $\Delta$Q for both, and a higher brightness for the Ba defect.

The Ba defect has a large formation energy, about 20 eV, which is similar to the $\mathrm{Xe_CVac_C}$.
Large formation energy is not necessarily a problem since XeV is suggested to be the most probable defect after Xe implantation~\cite{SANDSTROM2018182}.
ADAQ supports this conclusion since the single Xe substitutional has a higher formation energy of around 29 eV.

There are also more exotic complexes with a carbon interstitial and substitutional (I and Br).
Here, the Br defect has better properties than the NV center.
However, it also has a large formation energy of around 25 eV.
Furthermore, the defect with a K interstitial and substitutional has a huge formation energy of 43 eV, making it highly unlikely to form.
There is one substitutional-substitutional cluster with Mg with interesting ZPL that could be in the telecom range but also large formation energy, about 20 eV.
Furthermore, forming a double Mg substitutional cluster is extremely rare since substitutionals usually are stable with high energy barriers.
Hence, to form a cluster with two substitutionals by diffusion appear to be highly unlikely.

Finally, the search finds four single substitionals (H, F, K, and Xe).
The H and F dopants are offset from their ideal position, so they are more like $\mathrm{Int_H+Vac_C}$ and $\mathrm{Int_F+Vac_C}$, which they are named in ADAQ.
The K and Xe dopants sit at the ideal position with distortion of the nearest atoms depending on the dopant size.
There are two charge states for the K substitutional that have spin-1.

\begin{table}[h!]
\caption{Spin-1 defects on defect hull with ZPL larger than 0.5 eV and TDM larger than 3 debye.}
\begin{tabular} {cc|r|rrr}
Defect & Defect & Charge & ZPL & TDM & $\Delta$Q \\
Type & & & [eV] & [debye] & [amu$^{1/2}$\AA] \\
\hline
\multirow{5}{*}{\shortstack{XV\\known}}
 & $\mathrm{N_CVac_C}$ &  -1  &  1.70  &  6.66  &  0.56  \\
 & $\mathrm{Al_CVac_C}$  &  -1  &  1.0  &  5.92  &  0.51  \\
 & $\mathrm{Ga_CVac_C}$  &  -1  &  1.72  &  5.33  &  0.48  \\
 & $\mathrm{In_CVac_C}$  &  -1  &  1.88  &  6.31  &  0.33  \\
 & $\mathrm{Tl_CVac_C}$  &  -1  &  2.31  &  3.61  &  0.46  \\

\hline
\multirow{2}{*}{XV}
 & $\mathrm{Li_CVac_C}$  &  1  &  0.64  &  4.38  &  0.41  \\
 & $\mathrm{Ba_CVac_C}$ &  0  &  1.29  &  11.17  &  0.42  \\

\hline
\multirow{2}{*}{$\mathrm{X_CInt_C}$}
 & $\mathrm{I_CInt_C}$  &  -1  &  0.57  &  3.34  &  0.80  \\
 & $\mathrm{Br_CInt_C}$  &  -1  &  0.63  &  7.91  &  0.51  \\

\hline
\multirow{1}{*}{$\mathrm{X_CInt_X}$}
& $\mathrm{K_CInt_K}$  &  0  &  0.82  &  7.62  &  0.78  \\

\hline
\multirow{1}{*}{$\mathrm{X_CX_C}$}
 & $\mathrm{Mg_CMg_C}$  &  0  &  0.63  &  3.55  &  0.58  \\

\hline
\multirow{4}{*}{$\mathrm{X_C}$}
 & $\mathrm{K_C}$  &  -1  &  0.6  &  3.77  &  0.96  \\
 & $\mathrm{K_C}$  &  1  &  0.78  &  5.74  &  0.24  \\
 & $\mathrm{Xe_C}$  &  0  &  0.73  &  5.78  &  0.49  \\
 & $\mathrm{F_C}$ &  -1  &  1.20  &  9.83  &  0.42  \\
 & $\mathrm{H_C}$  &  -1  &  1.29  &  6.86  &  0.60  \\
 
\end{tabular}
\label{tab:spin-1}
\end{table}

\subsection{Database Search: Spin-\sfrac{3}{2} Defects}
\label{sec:spin-3/2}

There are many other searches to divide and filter the data which may be relevant to consider.
Here, we demonstrate how to find spin-\sfrac{3}{2} defects.
Table~\ref{tab:spin-3/2} shows stable defects with spin-\sfrac{3}{2}, a ZPL larger than 0.1 eV, and no TDM requirement to include spin-forbidden transitions.
The group 1 substitutionals show a similar trend.
As one goes down in the rows in the periodic table, the ZPL decreases.
The $\mathrm{Cs_C}$ is not included, since the predicted ZPL is below the cutoff in ADAQ (0.4 eV) and thus not calculated.
The Na stands out with increased ZPL from the trend but also with a high TDM (7.41 debye) and a low $\Delta$Q (0.14 amu$^{1/2}$\AA).
The K and Rb substitutionals could have emissions in the telecom range.
Only Be from group 2 substitutionals have spin-\sfrac{3}{2}.

The $\mathrm{Na_CNa_C}$ cluster has a formation energy of around 22 eV.
The $\mathrm{X_CInt_X}$ clusters have large formation energy of about 25 and 63 eV for Na and Sb, respectively.
However, the divacancy ($\mathrm{Vac_CVac_C}$) has a low formation energy of around 9 eV.
Of these defects, the divacancy, which consists of near neighbors, shows the most promising properties: a ZPL that can be in the telecom range with a strong TDM and a low $\Delta$Q.
The spin-\sfrac{3}{2} ground state is also found in Ref.~\onlinecite{PhysRevB.89.075203}.
The divacancy is also suggested to be the W29 center~\cite{KIRUI19991569}.

\begin{table}[h!]
\caption{Spin-\sfrac{3}{2} defects on defect hull with ZPL larger than 0.1 eV.}
\begin{tabular} {cc|r|rrr}
Defect & Defect & Charge & ZPL & TDM & $\Delta$Q \\
Type & & & [eV] & [debye] & [amu$^{1/2}$\AA] \\
\hline
\multirow{5}{*}{$\mathrm{X_C}$}
 & $\mathrm{Li_C}$ &  0  &  1.11  &  4.17  &  0.21  \\
 & $\mathrm{Na_C}$  &  0  &  1.49  &  7.41  &  0.14  \\
 & $\mathrm{K_C}$  &  0  &  0.46  &  5.83  &  0.6  \\
 & $\mathrm{Rb_C}$  &  0  &  0.36 &  8.64  &  0.61  \\
 & $\mathrm{Be_C}$  &  1  &  0.41  &  2.18  &  0.38  \\

\hline
\multirow{1}{*}{$\mathrm{X_CX_C}$}
& $\mathrm{Na_CNa_C}$ &  1  &  0.38  &  12.2  &  0.28  \\

\hline
\multirow{2}{*}{$\mathrm{X_CInt_X}$}
& $\mathrm{Na_CInt_{Na}}$  &  1  &  1.08  &  8.61  &  0.17  \\
& $\mathrm{Sb_CInt_{Sb}}$  &  -1  &  0.39  &  0.6  &  0.57  \\

\hline
\multirow{1}{*}{$\mathrm{VV}$}
& $\mathrm{Vac_CVac_C}$  &  -1  &  0.56  &  8.91  &  0.16  \\

\end{tabular}
\label{tab:spin-3/2}
\end{table}

\subsection{Database Search: Spin-2 Defects}
\label{sec:spin-2}

Table~\ref{tab:spin-2} shows stable defects with spin-2, a ZPL larger than 0.1 eV, and no TDM requirement to include spin-forbidden transitions.
$\mathrm{Na_CVac_C}$ has a formation energy of around 12 eV and the $\mathrm{F_CVac_C}$ around 9 eV.
The $\mathrm{Na_CVac_C}$ shows a spin-forbidden transition, hence the low TDM.

\begin{table}[h!]
\caption{Spin-2 defects on defect hull with ZPL larger than 0.1 eV.}
\begin{tabular} {cc|r|rrr}
Defect & Defect & Charge & ZPL & TDM & $\Delta$Q \\
Type & & & [eV] & [debye] & [amu$^{1/2}$\AA] \\
\hline
\multirow{2}{*}{XV}
 & $\mathrm{Na_CVac_C}$  &  -1  &  1.45  &  0.05  &  0.38  \\
 & $\mathrm{F_CVac_C}$ &  -1  &  1.53  &  3.82  &  1.18  \\

\end{tabular}
\label{tab:spin-2}
\end{table}

\subsection{Database Search:\\Spin Defects With a High Debye-Waller Factor}
\label{sec:Qdefects}

The previous searches show how to find spin defects on the defect hull.
However, there could be defects where the spin states are close in energy but the high spin is not on the defect hull.
We include defects with an increased distance of 30 meV (thermal energy at room temperature) above the defect hull in the search.
This criterion was also applied for the MgV center to argue that a higher laying spin state (22 meV than the ground state) can be stabilized by the thermal energy~\cite{Pershin2021}.
We also look for defects with a high spin state (S $\ge$ 1) and a high Debye-Waller factor (approximated with a low $\Delta$Q$<$0.2) yielding 5 unique defects shown in Table~\ref{tab:hqe}.
Apart from the already discussed divacancy in Table~\ref{tab:spin-3/2} and the MgV(0) center (also excellent predicted properties like divacancy but the spin-0 state is more stable according to Ref.~\onlinecite{Pershin2021}), all other defects contain sodium.
The $\mathrm{Na_CInt_{Na}}$ and $\mathrm{Na_C}(0)$ are also present in Table~\ref{tab:spin-3/2}.
The two new defects are the $\mathrm{Na_CNa_C}$ and $\mathrm{Na_C}(-)$.

\begin{table}[h!]
\caption{Defects 30 meV above the defect hull and spin larger or equal to 1 with ZPL large than 0.1 eV.}
\begin{tabular} {c|rr|rrr}
Defect & Charge & Spin & ZPL & TDM & $\Delta$Q \\
& & & [eV] & [debye] & [amu$^{1/2}$\AA] \\
$\mathrm{Vac_CVac_C}$  &  -1 & 3/2  &  0.56  &  8.91  &  0.16  \\
$\mathrm{Mg_CVac_C}$   &  0  &  1.0  &  0.58  &  7.69  &  0.12  \\
$\mathrm{Na_CInt_{Na}}$  &  1 & 3/2 &  1.08  &  8.61  &  0.17  \\
$\mathrm{Na_CNa_C}$  & 0 & 1 &  0.47  &  5.37  &  0.15  \\
$\mathrm{Na_C}$ &  0  & 3/2 &  1.49  &  7.41  &  0.14  \\
$\mathrm{Na_C}$ &  -1  & 1 & 1.50  &  7.27  &  0.15  \\
\end{tabular}
\label{tab:hqe}
\end{table}

The $\mathrm{Na_CNa_C}$ has a formation energy of around 22 eV and is similar to $\mathrm{Mg_CMg_C}$ discussed above.
The double Na substitutional cluster as first nearest neighbors has a formation energy about 4 eV lower than that of the two Na substitutionals as second nearest neighbors.
Even if the binding energy is significant for these defects, substitutionals usually require high temperatures to diffuse (as discussed for the $\mathrm{Mg_CMg_C}$ defect above).
The simplest sodium defect is the sodium substitional, which has formation energy around 13 eV and a spin-1 ground state within 1 meV from the defect hull.
By looking at the spin defects in diamond, sodium defects stand out with their presence in most searches: Table~\ref{tab:spin-3/2}, Table~\ref{tab:spin-2}, and Table~\ref{tab:hqe}.
Of these defects, the sodium substitutional has a good ZPL and a TDM with the lowest $\Delta$Q found in any defect search.
The spin-\sfrac{3}{2} state is on the defect hull, and the spin-1 state is extremely close ($<$1 meV) to the defect hull.
The sodium vacancy is one of only two dopants with a predicted spin-2 ground state.
Hence, we choose to study these defects in detail with high-order theory.

\subsection{Sodium Substitutional \ch{Na_C}}
\label{sec:nac}

The Na dopants have previously been considered as donors in diamond~\cite{doi:10.1063/1.110462}.
However, no shallow levels have been reported~\cite{POPOVICI19951305} and little change in resistivity is seen between Na and Ne implantation in diamond~\cite{HUNN1993847}.
The donor properties have been attributed to Na as an interstitial.
However, theoretical studies have found the Na substitutional more stable than the interstitial~\cite{PhysRevB.76.155203,PhysRevB.75.075202,LOMBARDI20081349}.
The ADAQ results support this conclusion.
As far as we know, the Na substitutional has not been considered for quantum applications, but with multiple spin states for the different charge states, it is a promising addition to the point defects in diamond.
Below are the results using the PBE and HSE06 functional.

\begin{figure}[h!]
  \includegraphics[width=\columnwidth]{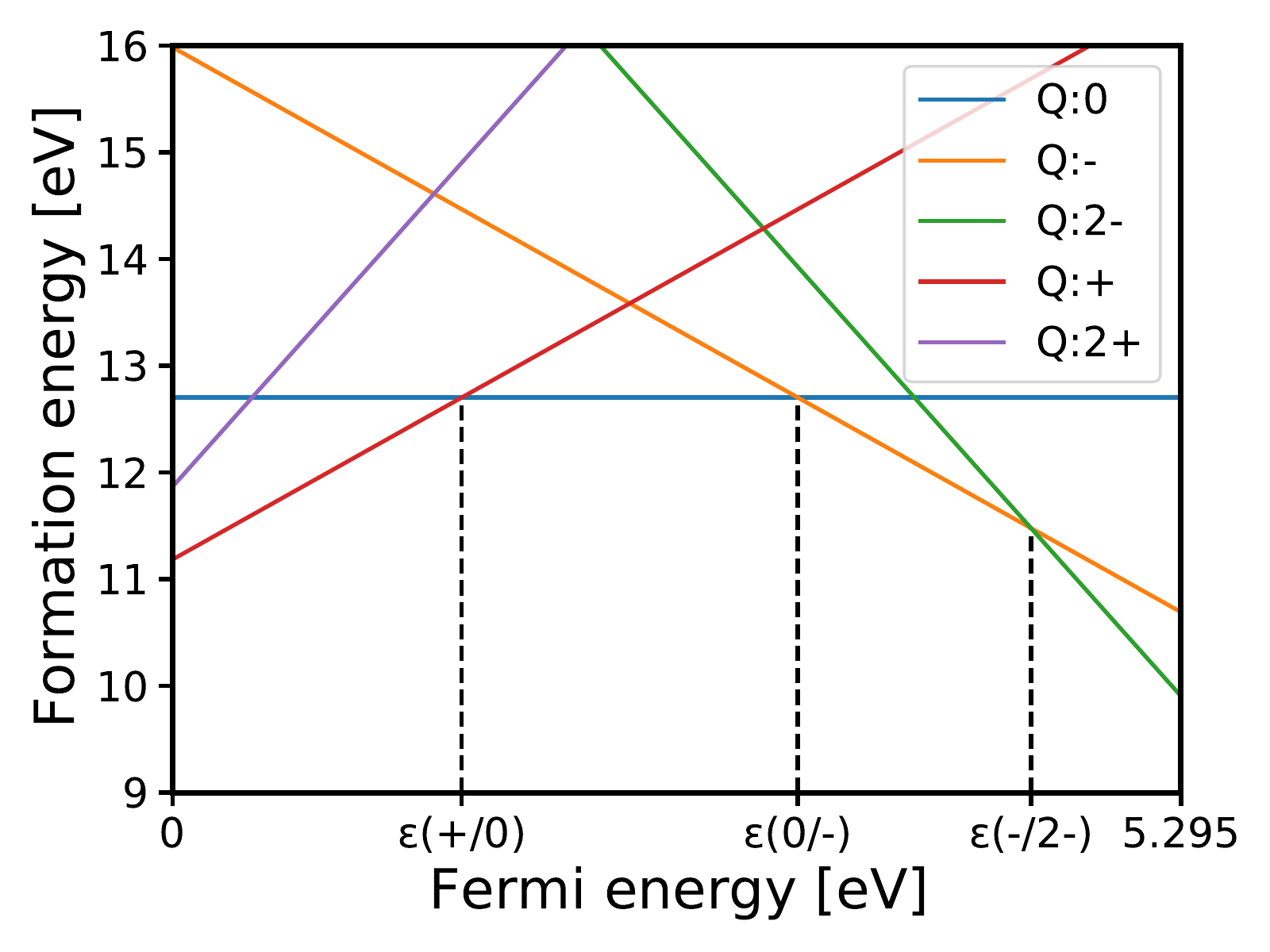}
	\caption{Formation energy of the different charge states (Q) of the sodium substitional as a function of the Fermi energy. The charge states are $\epsilon(+/0) = 1.520$ eV, $\epsilon(0/-) = 3.282$ eV, and $\epsilon(-/2-) = 4.507$ eV. The Fermi energy zero is set at the valance band maximum.}
	\label{fig:form} 
\end{figure}

Figure~\ref{fig:form} shows the formation energy for Na substitutional \ch{Na_C} calculated with the HSE functional.
ADAQ predicts that the negative charge state is spin-0 with an energy difference to the spin-1 state of about 1 meV.
This is contrasted with HSE results where the negative charge state with spin-1 is the lowest (95 meV lower in energy than spin-0) and stable in a wide region of Fermi energies.
Furthermore, the neutral charge state is spin-\sfrac{3}{2} with a wide stability region.
This charge state is found in ADAQ when searching for spin-\sfrac{3}{2} defects (see Table~\ref{tab:spin-3/2} with a predicted ZPL of 1.50 eV).
Furthermore, the positive charge state is also spin-1 but has no ZPL.
Hence, it did not appear in the ADAQ search (Table~\ref{tab:spin-1}).
The double positive charge state is not stable, but the double negative charge state is, with spin-\sfrac{1}{2} and a ZPL.

\begin{figure*}[t]
  \includegraphics[width=\textwidth]{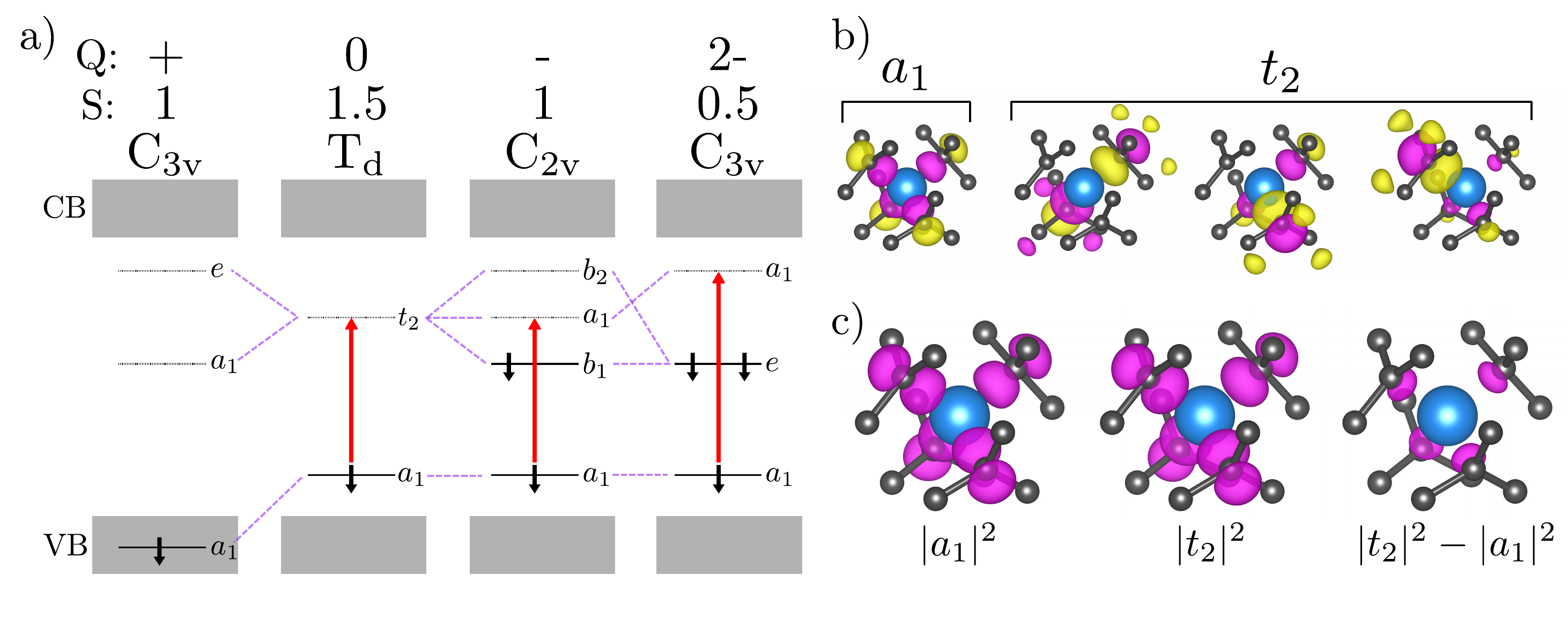}
	\caption{Electronic structure of the sodium substitional. a) Schematic eigenvalue level diagram of one spin channel for \ch{Na_C} in diamond, with charge state (Q), spin (S), and point group shown above each diagram. For more detailed image, including excited states, see supplementary material Figure S1. b) The defect orbitals of the neutral ground state calculated with the HSE functional. c) partial charge densities for the neutral ground state orbitals $|a_1|^2$,  the average of the three degenerate orbitals $|t_2|^2$, and difference $|t_2|^2-|a_1|^2$ with the same isosurface value as in Ref.~\cite{PhysRevApplied.11.044022} (0.025 \AA$^{-3}$). This isosurface only shows positive values for the partial charge density difference since the negative parts are smaller than the isolevel. C atoms are black, Na atoms are blue, purple (yellow) isosurfaces denote positive (negative) phase.
    }
	\label{fig:eigen} 
\end{figure*}

Figure~\ref{fig:eigen}a) shows the schematic eigenvalues for the stable ground states with marked ZPL transitions.
Except for the positive charge state, where the $a_1$ state moves below the valence band, all other states have similar ZPLs and TDMs, as reported in Table~\ref{tab:ZPL}.
The neutral charge state has the highest symmetry ($\mathrm{T_d}$) and two defect orbitals $a_1$ and $t_2$, see Figure~\ref{fig:eigen}b).
Figure~\ref{fig:eigen}c) shows the partial charge difference between these orbitals, which is much smaller than for the NV center and small compared to the silicon vacancy in SiC (cf. Figure 4 in Ref.~\onlinecite{PhysRevApplied.11.044022}).
These results suggest that \ch{Na_C} has a low spectral diffusion for the transition between these mid-gap states.
Similar transitions are observed in the negative and double negative charge states with only slight variations to the ZPL and TDM, attributed to Jahn-Teller splitting of the $t_2$ state when adding electrons.

\begin{table}[h!]
\caption{ZPLs, TDMs, and radiative lifetime $\tau$ for the optically active charge states of \ch{Na_C} calculated with the HSE functional.}
\begin{tabular} {c|cc|rrr}
Charge & Ground State & Excited State & ZPL & TDM & $\tau$ \\
State & Symmerty & Symmerty & [eV] & [debye] & [ns] \\
\hline
0  & $\mathrm{T_d}$    & $\mathrm{C_{2v}}$ & 1.592 & 6.8 & 13.5\\
-  & $\mathrm{C_{2v}}$ & $\mathrm{C_{3v}}$  & 1.682 & 6.6 & 12.1\\
2- & $\mathrm{C_{3v}}$ & $\mathrm{T_d}$ & 1.746 & 6.5 & 11.2\\
\end{tabular}
\label{tab:ZPL}
\end{table}

The \ch{Na_C} has multiple Jahn-Teller splittings for the different charge states.
For the neutral, there is a Jahn-Teller effect in the excited state; for the negative, in both the ground and excited states; for the double negative, in the ground state.
Table~\ref{tab:jahnteller} shows the Jahn-Teller stabilization energy ($E_{JT}=E_{min}-E_{T_d}$) calculated with the PBE and HSE functionals.
The PBE results show a low Jahn-Teller effect (around 10 meV) that increases to 50-80 meV with the HSE functional.
Compared to the NV center, the PBE functional gives a Jahn-Teller of 25 meV~\cite{PhysRevLett.107.146403} that increases to 42 meV with the HSE functional~\cite{PhysRevB.96.081115}.
To quantify the vibronic coupling, we use $\lambda = 2 \mathrm{E_{JT}} / n \hbar \omega$~\cite{bersuker_2006}, where $n$ is the degeneracy and $\hbar \omega$ is the zero-point energy of the vibration.
As a vibration estimate, we use the localized $t$ modes from the phonon calculations, see the supplementary material.
These have an energy of 21 meV for the PBE functional and 16 meV for the HSE functional.
With the PBE functional, $\lambda$ is about 0.25 for \ch{Na_C}, which is comparable to the NV center, whereas with the HSE functional, $\lambda$ is about 2-3, corresponding to a stronger vibronic coupling.
It is hard to state if the vibronic coupling is strong ($\lambda \gg 1$) or weak ($\lambda \ll 1$)~\cite{bersuker_2006}.
Hence, with the PBE functional, the NV center and \ch{Na_C} have similar Jahn-Teller effect.
However, with the HSE functional, the \ch{Na_C} becomes more static but still in an uncertain range.
Furthermore, the NV center is a $\mathrm{E} \otimes e$ Jahn-Teller effect, whereas the \ch{Na_C} is $\mathrm{T_2} \otimes (e+t_2)$~\cite{bersuker_2006,PhysRevMaterials.6.034601}.
Assuming only the $t$ phonons are involved (see supplementary material), it can be reduced to $\mathrm{T_2} \otimes t_2$.
However, to fully understand the Jahn-Teller effect in this system goes beyond the scope of this paper.
Such detailed analysis has been done for the \ch{Ni_C} (W8) defect~\cite{thiering2024spin}.

\begin{table}[h!]
\caption{The Jahn-Teller stabilization energy $E_{JT}$ in the ground or excited state marked with superscript $gr$ and $ex$, $\Delta$Q, and Debye-Waller calculated with one-phonon approximation and full phonons for the optically active charge states of \ch{Na_C}. The values inside the parenthesis show the Debye-Waller factor calculated without including the Jahn-Taller relaxation.}
\begin{tabular} {c|rr|rr|rr|rr}
Charge & \multicolumn{2}{c|}{$E_{JT}$} & \multicolumn{2}{c|}{$\Delta$Q} & \multicolumn{4}{c}{Debye-Waller [\%]} \\
State & \multicolumn{2}{c|}{[meV]} & \multicolumn{2}{c|}{[amu$^{1/2}$\AA]} & \multicolumn{2}{c|}{one phonon} & \multicolumn{2}{c}{full phonon}   \\
& PBE & HSE & PBE & HSE & PBE & HSE & PBE & HSE \\
\hline
0  & \makecell{$^{ex}$9} & \makecell{$^{ex}$54} & 0.17 & 0.52 & 69 & 17 & 77(85) & 40(83) \\
-  & \makecell{$^{gr}$9\\$^{ex}$7} & \makecell{$^{gr}$79\\$^{ex}$55} & 0.17 & 0.69 & 67 & 5 & $^\ast$75(85) & $^\ast$41(83) \\
2- & \makecell{$^{gr}$8} & \makecell{$^{gr}$67} & 0.17 & 0.63 & 68 & 6 & $^\ast$77(84) & $^\ast$27(83) \\
\end{tabular}
\label{tab:jahnteller}

\smallskip\footnotesize\centering 
$\ast$calculated with neutral charge state phonons.
\end{table}

\begin{figure}[h!]
  \includegraphics[width=\columnwidth]{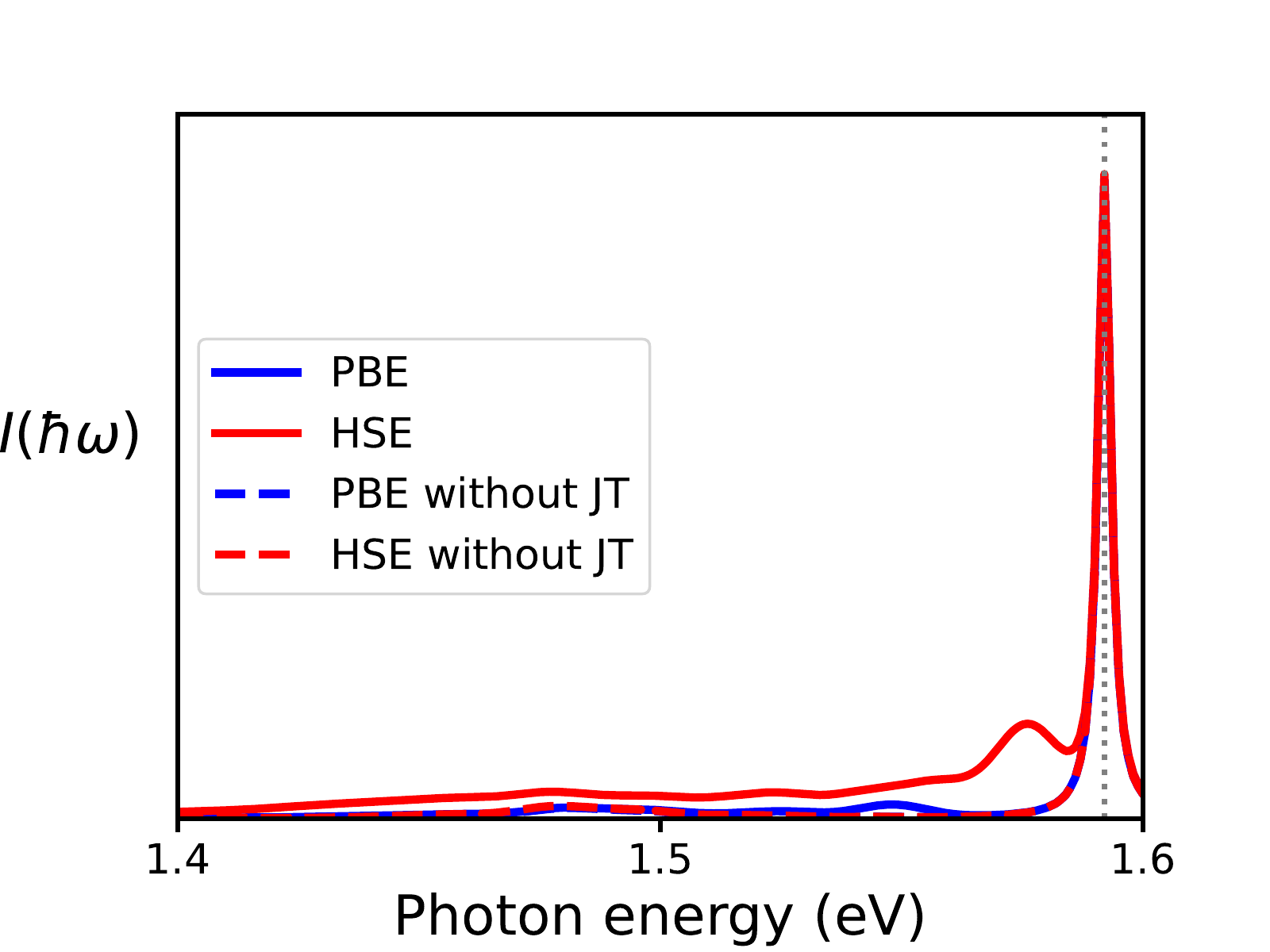}
	\caption{Calculated photoluminescence spectra for neutral charged \ch{Na_C} with the PBE and HSE functional. All spectra are shifted to align the ZPL to 1.592 eV. Including the Jahn-Teller effect, the Debye-Waller factor is 77\% and 40\% with the PBE and HSE functional, respectively. The first vibrational shoulder moves towards the ZPL with the HSE functional compared to the PBE functional, due to the strong Jahn-Teller effect. Excluding the Jahn-Teller effect, the Debye-Waller factor increases to 83\% for the two functionals with only minor differences in spectral shape.}
	\label{fig:spectra} 
\end{figure}

The \ch{Na_C} was predicted to have the lowest $\Delta$Q of all defect searches in the ADAQ database (see Table~\ref{tab:spin-1}, \ref{tab:spin-3/2}, \ref{tab:spin-2}, and \ref{tab:hqe}).
This continues to be the case for the more converged PBE results in all charge states, see Table~\ref{tab:jahnteller}.
However, with the HSE functional, the Jahn-Teller effect is much larger, and thus the $\Delta$Q increases.
Table~\ref{tab:jahnteller} also shows the Debye-Waller factors calculated with the one phonon approximation and for the full phonon calculation.
Here, the one phonon approximation underestimates the Debye-Waller factor compared with the full phonon calculation in all cases regardless of the Jahn-Teller effect.
In Figure~\ref{fig:spectra}, we show the difference in the calculated photoluminescence spectra with and without the Jahn-Teller relaxation for the neutral charge state with PBE and HSE functionals.
Due to the large Jahn-Teller relaxation for the HSE functional (see Table~\ref{tab:jahnteller}), the Debye-Waller factor is reduced from 77\% with the PBE functional to 40\% in the neutral charge state.
However, excluding the Jahn-Teller effect, the Debye-Waller factor is about 84\% for both functionals and all charge states.
In Ref.~\onlinecite{Alkauskas_2014}, the Jahn-Teller effect is neglected, which is a good approximation for the NV center.
However, this may not hold for the \ch{Na_C} due to the stronger vibronic coupling when using the HSE functional.
Regardless, the Debye-Waller factors are larger than the NV center.

Zero field splitting (ZFS) exists for the charge states with spin larger or equal to one, see Table~\ref{tab:ZFS}.
The neutral charge state has $\mathrm{T_d}$ symmetry, and therefore, the ZFS tensor is zero.
The negative charge state has $\mathrm{D_{2d}}$ symmetry with the PBE functional, hence there is no E splitting due to axial symmetry.
Whereas, calculations using the HSE functional gives $\mathrm{C_{2v}}$ symmetry and hence both a D and E splitting.
The PBE D result is in the same range as the D splitting of the silicon vacancy in SiC~\cite{vacancypaper}.
The ground state eigenvalues split due to crystal field splitting for the silicon vacancy in SiC and Jahn-Teller for the \ch{Na_C} in diamond.
However, the Jahn-Teller is much larger with the HSE functional, and consequently, the D splitting increases about 26 times, see Table~\ref{tab:ZFS} for values.
The positive charge state has $\mathrm{C_{3v}}$ symmetry with the HSE functional, thus only a D since E is zero.

\begin{table}[h!]
\caption{ZFS for the ground states of \ch{Na_C}.}
\begin{tabular} {c|cc|cc|cc}
Charge & \multicolumn{2}{c|}{Symmetry} & \multicolumn{2}{c|}{D [MHz]} & \multicolumn{2}{c}{E [MHz]}   \\
State & PBE & HSE & PBE & HSE & PBE & HSE \\
\hline
+ & $\mathrm{C_{2v}}$ & $\mathrm{C_{3v}}$ & -2787 & 6739 & -613 & $^\ast$0 \\
- & $\mathrm{D_{2d}}$ & $\mathrm{C_{2v}}$ & -120 & -3178 & $^\ast$0 & -857 \\
\end{tabular}
\label{tab:ZFS}

\smallskip\footnotesize\centering 
$\ast$due to symmetry.
\end{table}

To summarize, the \ch{Na_C} in diamond is a point defect that combines many sought-after properties for quantum applications.
\ch{Na_C} has an abundance of spin states.
The positive and negative charge states are spin-1, the neutral charge is spin-\sfrac{3}{2}, and the double negative is spin-\sfrac{1}{2}.
The neutral, negative, and double negative charge states all have a ZPL in the near-infrared region.
The ZPL, which is surprisingly close to the ZPL of the neutral single carbon vacancy (GR1)~\cite{Luhmann_2018,zaitsev2013optical}, is stable across the different charge states, meaning easy verification after implantation (and annealing) since the Fermi level does not need to be precisely controlled.
The Debye-Waller factor is predicted to be large, in fact, it is the largest found in the ADAQ database based on the $\Delta$Q.
If the Jahn-Teller effect is excluded, this defect has a Debye-Waller factor of about 84\% for all charge states regardless of the functional.
However, it is unclear how strong the vibronic coupling is since it depends on the functional used.
The HSE results likely provide a more accurate representation of intermediate vibronic coupling, neither weak nor strong.
Hence, the following are the most likely results:
A Jahn-Teller effect that results in a Debye-Waller factor as high as 40\% (for the neutral charge state), which is still good and far higher than the NV center at 3\%.
Further work is needed to determine the quantum yield and brightness based on rates of nonradiative decay and intersystem crossing~\cite{turiansky2024rational,doi:10.1021/acs.nanolett.2c04608}.
A ZFS with D values are in the GHz range for the negative charge state.
A low spectral diffusion as seen in Figure~\ref{fig:eigen}c) which is likely lower than the NV center.
However, experiments are needed to verify the strength of the Jahn-Teller effect and vibronic coupling as well as consequent effects on the Debye-Waller factor, ZFS, and spectral diffusion.

Let us discuss the neutral charge state of \ch{Na_C} in more detail due to only one Jahn-Teller effect in the excited state and the unique properties of spin-\sfrac{3}{2}.
There are many similarities between the neutral \ch{Na_C} in diamond and the negatively charged \ch{Vac_{Si}} in SiC.
Both defects have mid-gap states, are spin-\sfrac{3}{2}, and emit in the near-infrared region, specifically the first biological window~\cite{Weissleder2001}.
However, the Debye-Waller factor for \ch{Na_C} in diamond is much larger 
(40\% with Jahn Teller, 83\% without Jahn Teller) than that of the \ch{Vac_{Si}} in SiC (8-9\%~\cite{PhysRevApplied.13.054017}).
However, this does not include structural engineering, which increases the Debye-Waller factor to 58\% in nanowire~\cite{Lee2021} for the \ch{Vac_{Si}} is SiC.
The TDMs are comparable between the two defects (about 7 for \ch{Na_C} and 8 for \ch{Vac_{Si}}~\cite{ADAQ}).
For spectral diffusion, \ch{Vac_{Si}} in SiC has shown to have a small difference between the defect states (cf. Figure 4 in Ref.~\onlinecite{PhysRevApplied.11.044022}), whereas the \ch{Na_C} in diamond is even smaller, see Figure~\ref{fig:eigen}c).
The ZFS is small for the \ch{Vac_{Si}} is SiC, whereas it is zero for \ch{Na_C} in the ground state.
However, in the excited state, the ZFS will be nonzero due to symmetry breaking.
The spin-\sfrac{3}{2} and close to $\mathrm{T_d}$ symmetry make both defects excellent for strain sensing~\cite{PhysRevB.95.081405}.
To conclude, the spin-\sfrac{3}{2}, ZPL in the near-infrared (1.592 eV = 779 nm), strong TDM, possible stability against spectral diffusion, consisting of non-toxic elements, and the large Debye-Waller factor makes \ch{Na_C} in diamond ideal for biological quantum sensing.

\subsection{Sodium Vacancy (NaV)}
\label{sec:nav}

When implanting Na into diamond, apart from creating \ch{Na_C}, one also creates NaV centers (split-vacancy configuration).
In the ADAQ database, the NaV center has a slightly lower formation energy than \ch{Na_C}.
The formation energy trend is similar to the MgV center.
Since Mg is next to Na in the periodic table, we expect comparable yields during implantation (MgV about 36\% and single substitutionals \ch{Mg_C} about 15\%~\cite{Corte2023}).
A slightly higher yield of \ch{Na_C} may be due to the lower vacancy creation during implantation~\cite{Luhmann_2018}.
But, several NaV centers would still be created.
Let us briefly discuss their properties.
In the negative charge state, the NaV was predicted to be spin-2 by the database.
With the PBE functional, the spin-2 state is 409 meV lower than spin-1 and 631 meV lower than spin-0.
With the HSE functional, the spin-2 state is 283 meV lower than spin-1.
However, the spin-0 state is 29 meV lower than the spin-2.
A similar energy difference between the doublet and quartet is seen for the MgV center, where thermal energy at room temperature and strain could be enough to stabilize one spin state~\cite{Pershin2021}.

For the spin-2 state, the database predicts a ZPL of 1.45 eV with a forbidden transition, deduced from the minuscule transition dipole moment, see Table~\ref{tab:spin-2}.
Whereas, for the allowed transition between $a_{1g}$ and $e_u$ states, the HSE result gives a ZPL of 2.548 eV.
With the $\mathrm{D_{3d}}$ symmetry, this ZPL is stable against spectral diffusion.
Overall, the NaV center is quite similar to the MgV that has a ZPL of around 2.224 eV~\cite{Corte2023}.
However, the higher spin ground state (spin-2), compared with MgV spin-\sfrac{1}{2}, makes it a more interesting defect for spin control.
The spin-2 ground state could give rise to extended possibilities in spin control, as spin-\sfrac{3}{2} has for the silicon vacancy in SiC~\cite{PhysRevB.95.081405}.
To the best of our knowledge, this is the first prediction of a spin-2 point defect among all stable ground states for diamond defects.
The database shows only two defects with spin-2 ground state (see Table~\ref{tab:spin-2}).

\section{Discussion}

We start by discussing the search criteria used to reduce the number of interesting defects.
Overall, the PBE results, such as formation energy, spin, and ZPLs are accurate enough to identify relevant defects with interesting properties.
This paper demonstrates that all previously known defects are found on the PBE defect hull (see Table~\ref{tab:spin-1}).
However, some spin-1 defects close to the defect hull could be metastable or change spin state with more accurate methods, as was the case for the Na substitutional.
The conclusion is that the defect candidates found in the high-throughput screening with the PBE functional must be verified with higher-order methods.

For spin-1 defects, we also excluded defects with a TDM below 3.
These defects could either have a tiny TDM or a symmetry-forbidden transition.
ADAQ only calculates one excitation, and for some defects, a larger excitation could be allowed, like in the case of NaV.
Current automatic symmetry analysis efforts are being developed~\cite{Stenlund_msc_thesis,adaq-sym} to study these defects further.

The main focus of this paper has been on Na-related defects.
However, based on other results of the various searches presented in this paper, \ch{K_C} defects and F defects may also be relevant for further exploration.
The potassium substitutional \ch{K_C} also shows good ZPL, TDM, and $\Delta$Q.
The lower predicted ZPL could be in the telecom region but due to the size of the dopant, it was excluded since it will likely create more vacancies upon implantation than Na.
Florine ions are small and, hence, will create a small number of vacancies.
With good ZPL, TDM, and $\Delta$Q for the single substitutional (Table~\ref{tab:spin-1}) and spin-2 ground state for the fluorine vacancy (Table~\ref{tab:spin-2}), these defects are also interesting.
Experimental F implantation has already shown color centers~\cite{DitaliaTchernij2020}.
We encourage other researchers to explore the database themselves for additional interesting point defects in diamond.

The \ch{Na_C} was chosen for further study due to the lowest $\Delta$Q predicted by the high-throughput screening.
It turned out that for this defect, the Jahn-Teller effect changes between the ground and excited state when using a more accurate functional.
This change was unforeseen since this usually does not happen for most other defects.
For example, for the NV center, the Jahn-Teller values change a bit between functionals but do not show a quantitatively different behavior as in the case of \ch{Na_C}.
To improve the prediction of $\Delta$Q, one needs better geometry.
Currently, ADAQ has not implemented relaxation using the HSE functional in high-throughput.
For other defects, we assume that the $\Delta$Q trend is the same across all defects.
In other words, the PBE result underestimates the Jahn-Teller effect, and going to more accurate methods will keep or increase $\Delta$Q.
Even if the \ch{Na_C} did not have the predicted Debye-Waller factor.
It is still a previously unstudied quantum defect found by high-throughput.
Hence, the screening data provides relevant point defects of quantum applications.

To conclude, we have performed high-throughput calculations for extrinsic (s- and p-elements) dopants in diamond and collected the results in the ADAQ database.
The database contains not only known defects but also uncovers unexplored defects for quantum applications, such as sodium.
Implantation of Na into diamond would yield an array of interesting spin defects for quantum applications.
The Na single substitutional has spin-1 in the negative charge state, a bright ZPL in the near-infrared region, and a high Debye-Waller factor.
The neutral charge state has similar properties but with a spin-\sfrac{3}{2}.
Both charge states of \ch{Na_C} have properties that make them excellent for biological sensing applications, better than the NV center due to bright ZPL in the near-infrared, possibly lower spectral diffusion, and the much increased Debye-Waller factor.
Furthermore, the database also contains the Na vacancy cluster, one of two defects predicted to have a spin-2 ground state.
These findings of sodium-related defects with high spin states show the usefulness of high-throughput screening.
This work presents the ADAQ database, a powerful tool for finding stable point defects relevant to quantum applications.

\section*{Methods}
\label{sec:methodology}

\subsection*{High-Throughput calculations}
The high-throughput calculations were performed with the software package ADAQ~\cite{ADAQ,adaq_info}, which, in turn, carries out automated workflows implemented using the high-throughput toolkit (\emph{httk})~\cite{armientoDatabaseDrivenHighThroughputCalculations2020} with the density functional theory (DFT) calculations being preformed by the Vienna Ab initio Simulation Package (VASP)~\cite{VASP, VASP2}.
VASP uses the projector augmented-wave method~\cite{PAW, Kresse99}.
The exchange-correlation functional is the semi-local functional by Perdew, Burke, and Ernzerhof (PBE)~\cite{PBE}.
The defects are simulated at the gamma point in a $4\times4\times4$ cubic supercell containing 512 atoms, with a lattice constant of $3.57$ Å.

\subsection*{Hybrid calculations}
For the sodium defects, the functional is the HSE06 hybrid functional by Heyd, Scuseria, and Ernzerhof (HSE)~\cite{HSE, HSEerrata}.
Sodium was simulated with the pseudopotential with $2p^6 3s^1$ electron configuration (Na\_pv), and carbon was simulated with the pseudopotential with $2s^2 2p^2$ (C).
The ionic and electronic stopping parameters are $5\cdot10^{-7}$ eV and $1\cdot10^{-8}$ eV respectively, and the cutoff energy of the plane-wave basis set is $600$ eV.
Phonopy~\cite{phonopy, phonopy-phono3py-JPSJ} is used to calculate the phonons in the neutral ground state with both PBE and HSE, to speed up this large set of calculations the electronic stopping parameter is $1\cdot10^{-6}$ eV, and the plane-wave cutoff energy is $400$ eV.
The excited states are simulated by promoting one electron to occupy a higher orbital and constraining this configuration when relaxing the electronic structure~\cite{PhysRevLett.103.186404}, the ZPL is the total energy difference between the relaxed ground and excited states. 
Symmetry constraints are not used when relaxing the crystal structure, this allows Jahn-Teller distortions to occur.
When running simulations to calculate $\mathrm{E_{T_d}}$, the $\mathrm{T_d}$ symmetry is fixed during the ion relaxation.
The photoluminescence spectra and Debye-Waller factor are computed using Pyphotonics~\cite{Tawfik2022}.
The neutral charge state phonons are assumed to be similar to the other charge state phonons and hence used to estimate the Debye-Waller factor.
The ZFS tensor is calculated with the VASP implementation of the method by Ivády et al.~\cite{PhysRevB.90.235205}.
The point group symmetry of the crystal structure is found with AFLOW-SYM \cite{Hicks:AFLOW-SYM}.
Defect orbital symmetry analysis and polarization selection rules of optical transitions is done with the method in Ref.~\onlinecite{adaq-sym}.
TDM~\cite{davidsson2020} is calculated with a modified version of PyVaspwfc~\cite{QijingZheng}.
To obtain the TDM of a transition from a degenerate state, the average is taken of the element-wise absolute value of each orbital in the degenerate state.
Radiative lifetime is calculated by taking the inverse of the Einstein coefficient~\cite{davidsson2020}, with the refractive index of diamond $2.4$.
The isosurface level used in Figure~\ref{fig:eigen} c) is $0.025 \mathrm{Å^{-3}}$, same as Ref.~\onlinecite{PhysRevApplied.11.044022}, while the isolevel in b) is $\sqrt{0.025} \mathrm{Å^{-3/2}} = 0.1581 \mathrm{Å^{-3/2}}$, meaning the $\mathrm{a_1}$ orbital in b) has the exact same shape as $|a_1|^2$.

\section*{Acknowledgments}

We acknowledge support from the Knut and Alice Wallenberg Foundation (Grant No. 2018.0071).
Support from the Swedish Government Strategic Research Area Swedish e-science Research Centre (SeRC) and the Swedish Government Strategic Research Area in Materials Science on Functional Materials at Linköping University (Faculty Grant SFO-Mat-LiU No. 2009 00971) are gratefully acknowledged.
This work was partially supported by the Knut and Alice Wallenberg Foundation through the Wallenberg Centre for Quantum Technology (WACQT).
JD and RA acknowledge support from the Swedish Research Council (VR) Grant No. 2022-00276 and 2020-05402, respectively.
The computations were enabled by resources provided by the National Academic Infrastructure for Supercomputing in Sweden (NAISS) and the Swedish National Infrastructure for Computing (SNIC) at NSC, partially funded by the Swedish Research Council through grant agreements no. 2022-06725 and no. 2018-05973.

\section*{Data Availability}

For more information about the ADAQ database, see \url{https://httk.org/adaq/}.

\section*{Code Availability}

For more information about the ADAQ source code, see \url{https://httk.org/adaq/}.

\section*{Competing interests}
The authors declare no competing interests.

\section*{Author contributions}
J.D. conceptualized the project in discussion with R.A., analyzed the data, and wrote the manuscript.
W.S. performed the sodium defects calculations and made the figures.
A.P. performed the high-throughput calculations.
R.A. and I.A.A. supervised the project and reviewed the manuscript.


\bibliography{references}

\end{document}


\title{Supplementary Material for Na in Diamond: High Spin Defects Revealed by the ADAQ High-Throughput Computational Database}

\author{Joel Davidsson}
\email{joel.davidsson@liu.se}
\affiliation{Department of Physics, Chemistry and Biology, Link\"oping University, Link\"oping, Sweden}

\author{William Stenlund}
\affiliation{Department of Physics, Chemistry and Biology, Link\"oping
 University, Link\"oping, Sweden}

\author{Abhijith S Parackal}
\affiliation{Department of Physics, Chemistry and Biology, Link\"oping University, Link\"oping, Sweden}

\author{Rickard Armiento}
\affiliation{Department of Physics, Chemistry and Biology, Link\"oping University, Link\"oping, Sweden}

\author{Igor A. Abrikosov}
\affiliation{Department of Physics, Chemistry and Biology, Link\"oping University, Link\"oping, Sweden}

\maketitle

\section{Detailed ground and excited state Kohn-Sham eigenvalues}

\begin{figure}[h!]
  \includegraphics[width=\textwidth]{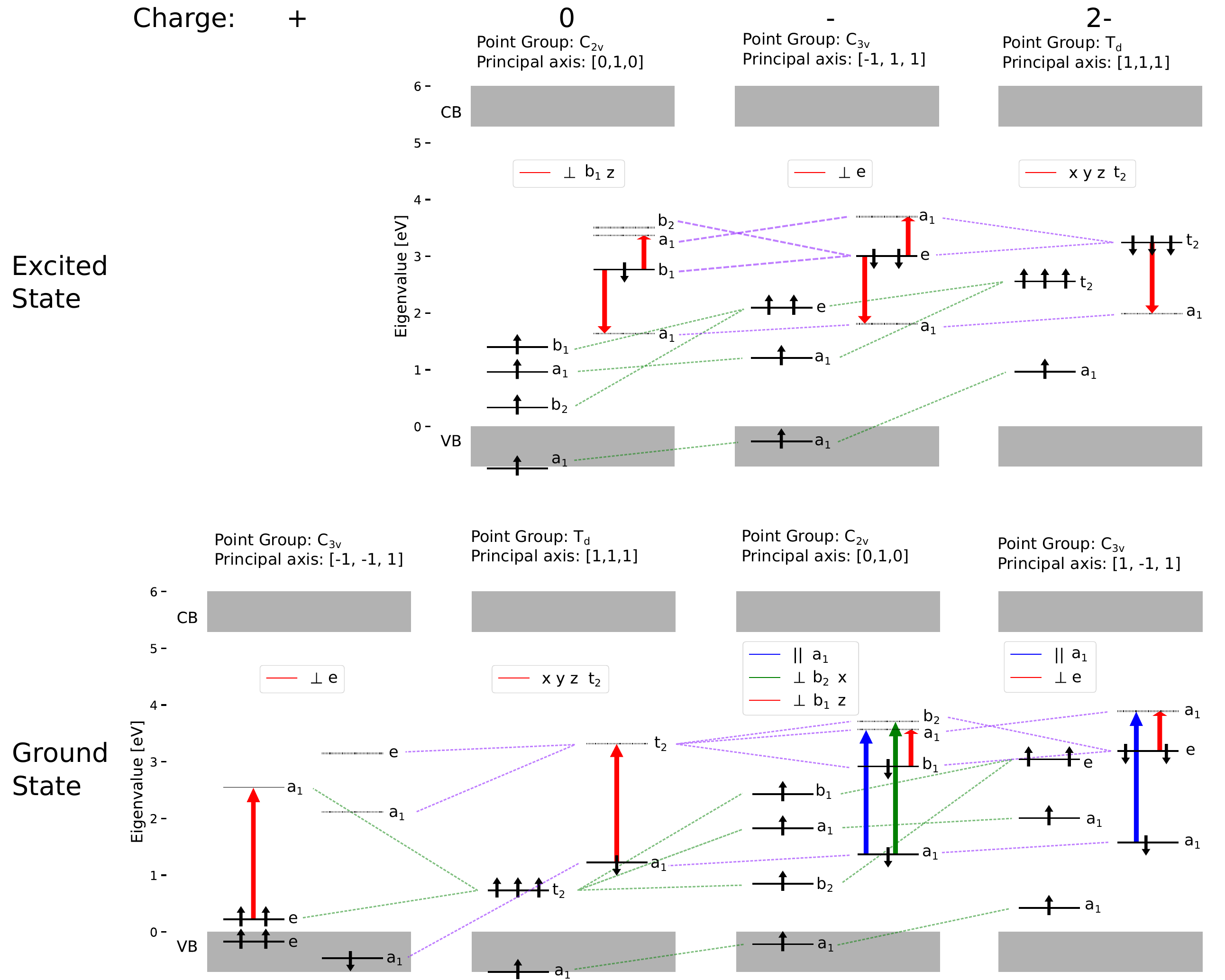}
	\caption{Eigenvalues of states in the band gap for ground and excited states in different charge states. Point group and principal axis is shown above each diagram. Occupied bands levels are solid and the spin is indicated with a black arrow, empty levels are dotted. The irrep of each level is marked, the selection rule allowed optical transitions are drawn as colored arrows, the legend in each diagram shows the irrep och each color and the polarization direction relative to the principal axis. Note that the principal axis given for $\mathrm{T_{d}}$ is [1,1,1], which is not unique, any of the four body diagonals are valid.}
	\label{fig:eigen_full} 
\end{figure}

Supplementary figure~\ref{fig:eigen_full} presents the \ch{Na_C} defect orbitals for all charge states as well as both ground and excited states.
The irreducible representations, the selection rules, and the diagrams were found and made with ADAQ-SYM~\cite{Stenlund_msc_thesis,adaq-sym}.

Note that there are Jahn-Teller distortions of the crystal structure between ground and excited states, which have different principal axes.
For example, the blue arrow in the negative ground state indicates that the lowest spin down electron can absorb light polarized parallel to the principal axis [0,1,0]
In contrast, the excited negative state emits light polarized perpendicular to a different principal axis [-1,1,1], this is because the crystal distorts due to a Jahn-Teller effect when in the excited state.

The HSE calculations show that when a triply degenerate state is empty or fully occupied, the symmetry remains as $\mathrm{T_d}$, with one electron in the degenerate state, a Jahn-Teller distortion relaxes the structure to a $\mathrm{C_{2v}}$ symmetry, and when there are two electrons in the degenerate state, the structure relaxes to a $\mathrm{C_{3v}}$ symmetry and the two electrons form a degenerate $e$ state.
For the $\mathrm{C_{2v}}$ distortion, the principal axis is one of the three Cartesian directions, and for the $\mathrm{C_{3v}}$ distortion, the principal axis is one of the four body diagonals of the cubic supercell.
If a $\mathrm{T_{d}}$ structure was relaxed with one or two electrons put in the degenerate $\mathrm{t_{2}}$, the structure may distort along different directions since there are three and four equivalent minima for the $\mathrm{C_{2v}}$ and $\mathrm{C_{3v}}$ respectively.

\newpage

\newpage
\section{Phonons for the neutral charge state \ch{Na_C}}

\begin{figure}[h!]
  \includegraphics[width=\textwidth]{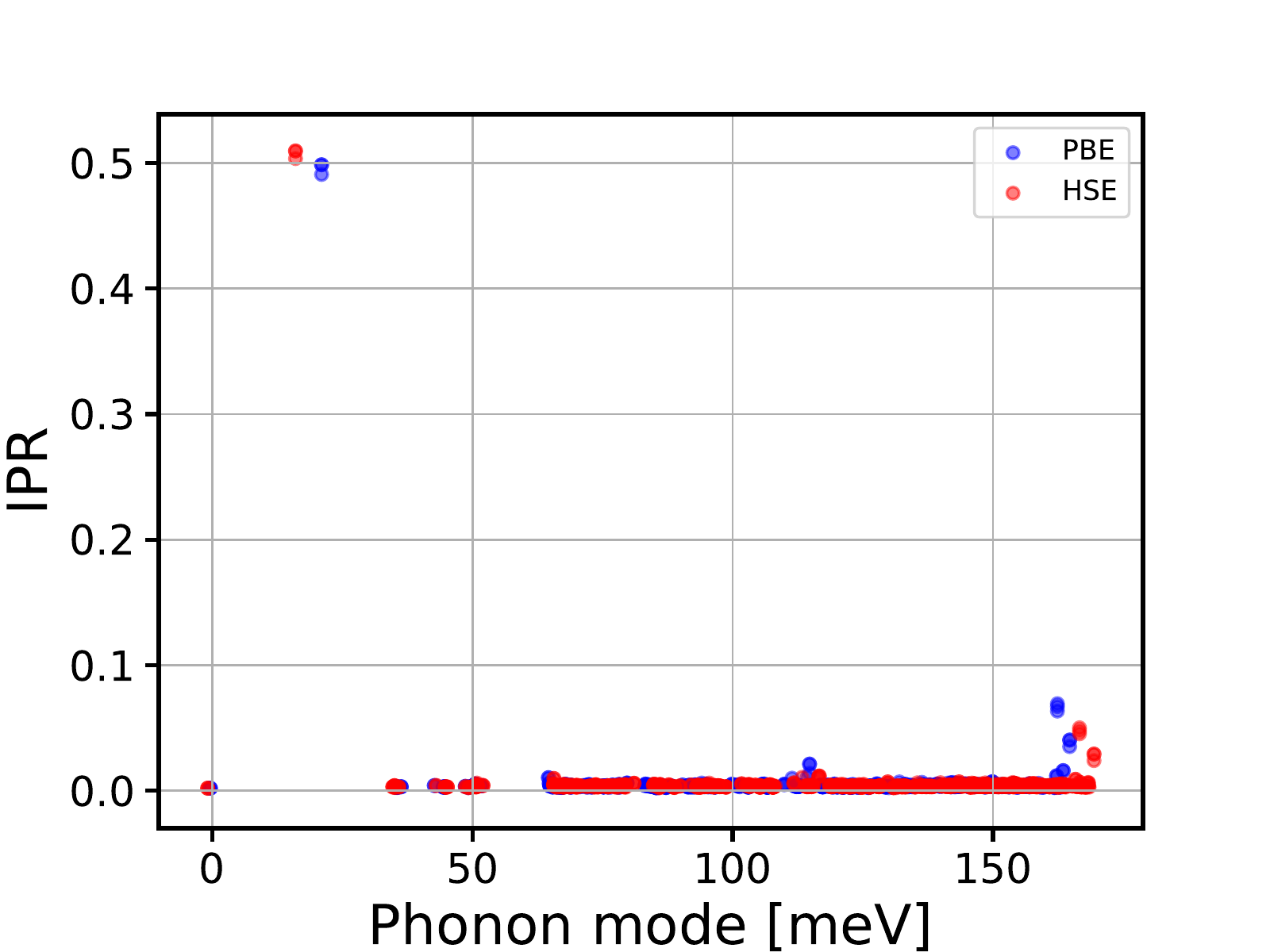}
	\caption{Phonon modes plotted with energy vs inverse participation ratio (IPR) for the \ch{Na_C} ($\mathrm{T_d}$ symmetry) defect with the PBE and HSE functional.}
	\label{fig:phonon} 
\end{figure}

Supplementary figure~\ref{fig:phonon} shows the phonon spectra for the \ch{Na_C} point defect in the neutral charge state.
There is a localized $t$ phonon mode at 21 meV and 16 meV for the PBE and HSE functional, respectively.
There are also lower localized $t$ phonon modes at 160-170 meV.

\bibliography{references}